# The Interplanetary Habitable Zone

Caleb Scharf, NASA Ames Research Center, Moffett Field, California, USA, caleb.a.scharf@nasa.gov

**Abstract:**

The concept of a system-wide measure of the sustainment of life (habitability) for space-faring interplanetary species is introduced and explored. Although largely agnostic to the details of how interplanetary life might operate (e.g., via technology or by utilizing organism traits that are, as of now, unknown to us), some assumptions must be made about energy harvesting, orbital mobility costs, radiation risks, and resource requirements. A multi-modal figure of merit is developed for evaluating an interplanetary habitable zone (IHZ). An agent-based model is also developed to simulate the dispersal of interplanetary life in a planetary system and characterize the IHZ. For the solar system, resource weightings between planetary bodies dictate many overall behaviors, including the sequence of migration from Earth to the Moon, Mars, and asteroid belt. Comparisons with the Trappist-1 exoplanetary system also point to critical sensitivities in the balance between resource availability and risk or cost factors (e.g., radiation risks and orbital Delta-v costs) that determine the structure of an IHZ. Results suggest that our solar system may have an inherent, and significant, advantage for a space-faring species over a system like Trappist-1. This modeling approach may also have application to emerging space economies in our own solar system.

*Submitted to The Astrobiology Journal 2/13/2026*

## 1. Introduction

Following works such as Kasting et al. (1993) (and historically, Newton, 1678; Maunder, 1913; Huang, 1966) the concept of a circumstellar habitable zone (CHZ) has been widely explored and elaborated to characterize rocky, terrestrial-type planets in terms of their surface environments and their potential for originating and sustaining life (with considerable complexity in the finer details, e.g., Cockell et al., 2016, Ramirez, 2018). That characterization is often presented as a binary "whole planet" choice, based on the orbital range calculated for the CHZ and the planet's location and potential atmospheric composition, and may include factors such as geophysical evolution or stellar activity. In other analyses a measure of spatial or temporal fractional habitability has been derived based on surface area or seasonal intervals that also incorporate more detailed models of planetary climate and influential factors such as spin-orbit properties or land-ocean arrangements (e.g., Spiegel et al. 2008; Jansen et al 2019).

In addition to providing a yardstick for studying the history of worlds in the solar system, the CHZ is a prominent metric for categorizing exoplanets based on their size, orbit, and stellar host (e.g., Kopparapu et al. 2013), and for prioritizing targets for more elaborate astronomical follow-up (e.g., to constrain atmospheric compositions). Extensions to the classical CHZ also include estimates of the equilibrium temperature of icy bodies to assess the likelihood of closed ocean worlds (planets or moons) as another class of habitable environment (e.g., Scharf 2006; see also Figure 1).

Underlying most applications of the CHZ is the assumption that life is confined to its point of planetary origin. Or, at most, has undergone some form of limited exchange or cross-contamination due to the transfer of impact ejecta between similarly habitable bodies (e.g., Scharf and Cronin 2016, Lingam and Loeb 2017), that simply serves to modify the probability of life originating on a given body.

However, this leaves open the question of habitability in systems where life has actively extended or expanded beyond its origins to utilize a far broader array of environments. In our own solar system, it could be argued that, via human space exploration, life is in the process of becoming interplanetary – both in terms of a spread in biological presence, and in terms of a co-evolutionary technological presence that could be considered a part of an extended phenotype of terrestrial life (c.f., Frank et al. 2022, Scharf 2025). In the future this could result in a much larger dispersal of Earth's life and its associated infrastructure (in technology and other engineered structures) to other natural bodies and to new, built habitats or functional structures (e.g., for computation, Scharf and Witkowski 2024). In such circumstances, the degree to which a system can sustain life will be determined by many factors in addition to those of the classical CHZ. Evaluating an Interplanetary Habitable Zone (IHZ) could therefore provide insight to life's opportunities in our own solar system, the plausibility of interplanetary life in exoplanetary systems (thereby also prioritizing systems for study of extended biosignatures or technosignatures), and help to evaluate the statistical potential across the galaxy of interplanetary life by mapping exoplanetary system properties to IHZs.

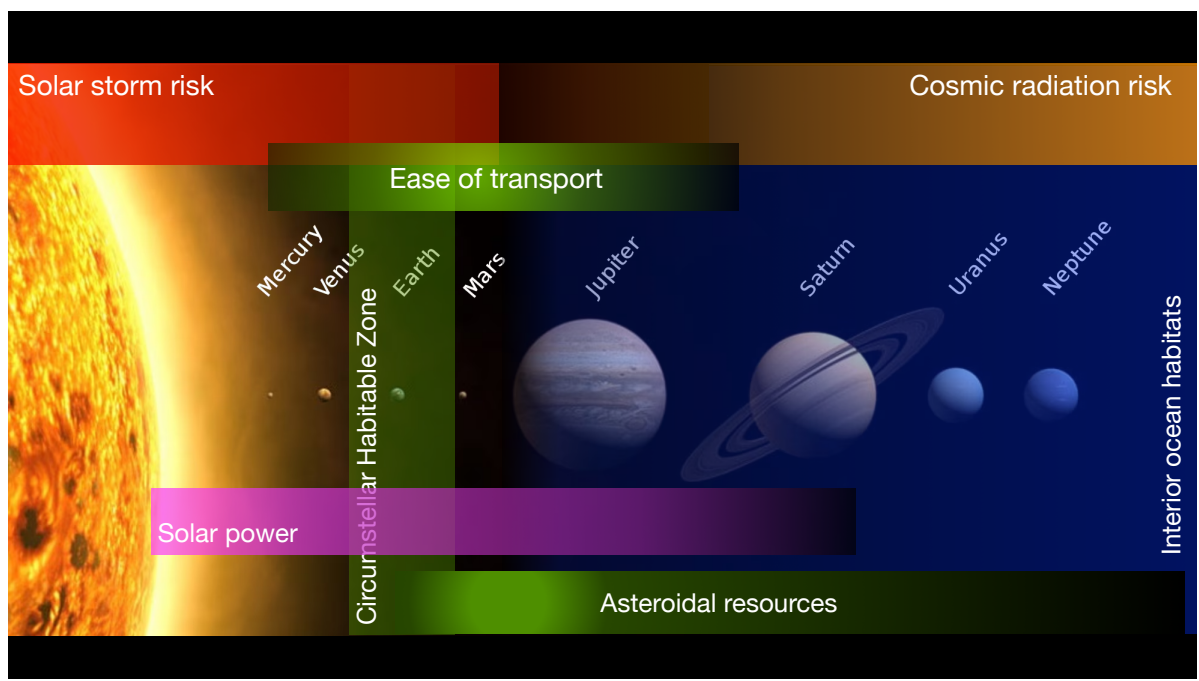

*Figure 1: Illustration of some of the multi-modal factors that may contribute to an IHZ in our solar system, overlaid on traditional habitable zones: the circumstellar habitable zone and the zone where closed icy ocean worlds can exist.*

In this paper an initial study is made of habitability for interplanetary life that focuses on some of the multi-modal factors that might define limits or opportunities for the sustainment of life that disperses from its origin point (Figure 1). In this approach the IHZ is a more explicitly multi-dimensional property than the CHZ, which is often reported just in terms of orbital range, even though it has multiple dependencies (e.g., planetary atmosphere, stellar parent etc.) In this study of the IHZ, many factors to do with the surface or subsurface suitability (habitability) of planetary bodies or other locations that sustain populations of dispersing, space-faring life are not identified specifically, but are folded into a resource weighting factor (see Discussion).

In Section 2 the factors under consideration are discussed in detail. In Section 3 an overall formalism for the IHZ is proposed and applied in a limited fashion to the solar system and the Trappist-1 exoplanetary system. In Section 4 an initial study is presented of an agent-based simulation that attempts to capture the dynamics of a species spreading from its point of origin in a more sophisticated way, to further explore the properties of the solar system's IHZ and that of the Trappist-1 system.

## 2. Factors contributing to an Interplanetary Habitable Zone (IHZ)

### 2.1 Power availability

Interplanetary life will require power whether in empty space or at any other location. The luminosity of main sequence stars is assumed to dominate the total available power budget of most planetary systems (i.e., exceeding engineered producers such as exothermic chemistry, nuclear fission reactors, or nuclear fusion reactors). The potential capture of stellar power (luminosity $L_*$) by finite size receivers will track the flux at a distance $r$: $f_* = L_*/(4\pi r^2)$, but is also determined by the efficiency of capture systems and their environmental dependencies, such as thermal conditions (given by an equilibrium temperature $T_{eq}$). For example, modern commercial photovoltaic panels on the Earth decline in efficiency of power output on average by 0.3-0.5% for every degree Celsius above ~25°C (~298 K). A useful form for the power efficiency ($\eta$) as a function of temperature is therefore a simple linear relationship relative to a reference temperature $T_{ref}$. I.e. $\eta = \eta_{ref}[1 - \beta(T_{eq} - T_{ref})]$, where the temperature coefficient $\beta$ (the % efficiency change per degree Celsius) might range between 0.3% and 0.5%.

To first order, assuming radiative equilibrium with normally incident stellar power, the mean temperature of a *flat* surface photovoltaic receiver will be $T_{eq} = [(1-A)f_*/\sigma]^{1/4}$ where $A$ is the albedo of the receiver and $\sigma$ is the Stefan-Boltzmann constant and re-radiation is assumed from one side only. Since $f_*$ is the stellar flux at a distance $r$ then $T_{eq} = [(1-A)L_*/(4\pi\sigma r^2)]^{1/4}$.

Figure 2 illustrates the resulting functional form for the solar power figure of merit: $P_* = \eta \cdot f_*$ as a function of distance from a star, with a 25°C reference temperature for solar power efficiency. This can be used as a first-order measure of the effective availability of solar power at any location in a system.

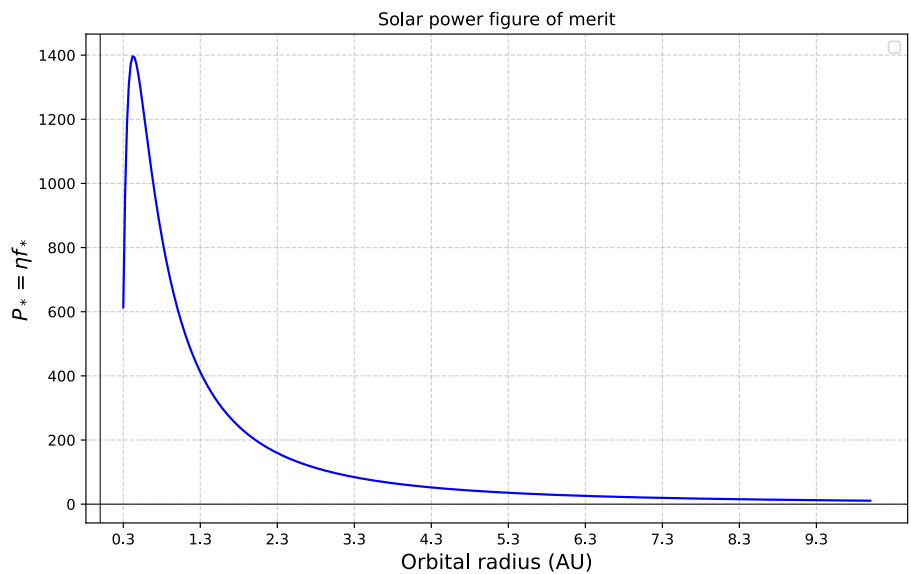

*Figure 2: The functional form of a figure of merit for solar power $P_* = \eta \cdot f_*$.*

## 2.2 Radiation risk

Interplanetary life and its engineered structures may be directly exposed to the solar wind and to higher energy solar energetic particles (SEPs) as well as cosmic radiation originating outside of a planetary system (galactic cosmic rays, GCRs). Particle radiation can cause a variety of detrimental effects, from biological damage to material structural damage, as well as erroneous operations in electrical devices (including digital computation). Although radiation exposure of biology or technology may be reduced within atmospheres or by shielding materials during interplanetary transit, or using materials such as regolith on a surface, these can add a sustainability cost and may not avoid a lifetime integrated risk that depends on location.

In our solar system, the general behavior of the mean solar wind particle rates follows an inverse-square law drop-off of particle density (primarily protons, electrons, and helium nuclei) with distance from the Sun. In addition, the average kinetic temperature of particles follows a $\sim 1/r^{0.57}$ relationship – although with significant variations in time and overall radial velocity. But these characteristics can be very significantly, albeit temporarily, affected by coronal mass ejections (CMEs) generating SEPS, and other solar activity (see e.g., Guo et al 2024).

The quiescent solar wind is considered relatively benign (with typical heavy particle energies <10 keV and electron energies <100 keV), while solar flare events and CMEs can inject higher energy, relativistic particles into the flow (with energies up to GeV). However, even quiescent solar wind can produce a variety of damages: e.g. microscopic blistering of metallic surfaces by H+ accumulation indicate deleterious effects after 1 year at 1AU (e.g., Sznajder et al. 2019 and references therein). This kind of erosive effect could damage solar sails, sun shields, and insulation/thermal coatings on space-based devices.

Solar CMEs have the highest risk potential for interplanetary life (for material damage or biological damage), with characteristics for particle output that are diverse and are not restricted to low ecliptic inclinations. These events are also strongly dependent on stellar cycles, e.g. the Sun's activity cycle, and can be correlated with stellar mass and age activity cycles on other sun-like stars e.g., Jeffers et al. 2023).

While SEPs present one high risk factor, GCRs present another with the potential for much higher particle energies (MeV to PeV). In our solar system, the structure and extent of the heliosphere play a significant role in modulating cosmic radiation levels and reducing risk across much of the orbital range of the major planets during higher solar activity periods. Stellar wind models have been used to estimate that cosmic radiation in exoplanetary systems, where faster rotating and more active younger stars can more effectively block cosmic radiation. Although this varies depending on particle energy, with shorter particle diffusion timescales occurring at higher energies. The proximity to planetary magnetic fields is also a factor in exposure levels to both SEPs and GCRs, whether on a planetary surface or in near orbit.

In the solar system the radial increase of GCR rates averaged over time has been estimated as <10% per AU within 0.5 AU of the Sun and this appears consistent for distances beyond 1 AU where radial gradients of 2% to 4% per AU have been estimated (see e.g., Lawrence et al. 2016 and references therein for GCR rates at 1AU to over 80AU.) All rates vary depending on factors such as solar modulation and heliocentric latitude.

Consequently, for a figure of merit of the radiation risks for the IHZ, the time-averaged risk (particle rates) might be approximated to first order as:

$$\lambda_{radiation}(r) = A_{GCR} \cdot r + A_{Solar} \cdot r^{-2}$$

Where $A_{GCR}$ is a normalization constant for cosmic radiation rates/risks that are assumed to increase linearly with distance $r$ from the host star, and $A_{Solar}$ is a normalization constant for solar radiation rates/risks that is assumed to decrease as $r^{-2}$. It should be noted that the exact drop-off with orbital radius is seen to vary with particle energy (e.g. SEP rates can sometimes more closely follow a $r^{-3.5}$ drop-off) and particle type (e.g. with electron rates also often following a steeper decline with radius) and transport factors such as interplanetary magnetic fields. Therefore, the inverse-square form suggested here is a conservative choice in terms of larger risks at larger orbital radii. There may be additional radiation risk factors to consider, such as a planetary magnetosphere (e.g., Jupiter), and a more detailed IHZ model could account for these (see Section 4 below).

**2.3 Orbital transfer and potential well costs**

The energy, or Delta-v (applied velocity change), requirements to "get around" in a system should play a significant role in determining the dynamics of interplanetary dispersal and the efficiencies, capacities, and timescales of transportation of resources. A full analysis of any system requires extensive modeling based on the properties of natural bodies and their orbits to determine the potential networks of movement and their classification. For example, fast hyperbolic transfers incur high Delta-v cost. While more energy efficient medium Delta-v transfer pathways (e.g. Hohmann-style elliptical transfers) involve somewhat slower transit times and are also more constrained in timing and total energy cost. They are therefore also more readily quantifiable in any model of the IHZ. Other transfer strategies include the bi-elliptic transfer, but this requires a high ratio (12:1) between orbital radii and might therefore have limited utility.

Gravity assists or weak stability boundary manifold pathways (e.g. Belbruno 1987) can involve low to ultra-low Delta-v budgets and could be favored for a species transporting non-perishable resources around a system. However, these kinds of trajectories are highly sensitive to the configuration of planets or other objects and require much more specialized and intensive computational work to assess for the IHZ. The same costs in timing and analysis would also exist for a species to implementing these kinds of transfers, potentially constraining their use.

An additional set of Delta-v costs come from any maneuvers to or from the surface of a body, that would be necessary to access most material resources (Figure 3). Although other forms of access to planetary surfaces, such as static structures (e.g., space elevators) could improve efficiencies, the total energy expenditure of transfer from surface to space remains significant. Hence, the surface escape/landing Delta-v budget across a planetary system is also a factor to use in evaluating the IHZ, with the Delta-v maximum envelope set by the full escape velocity (i.e., as opposed to low altitude orbits around bodies). It is noted that since escape velocity from an object's surface is: $v_e = \left(2GM_p/R_p\right)^{1/2}$, contributions of planetary rotation to the budget for on/off maneuvers from a planetary surface are unlikely to be significant. For the Earth, for example, low-latitude spinward launches only gain ~0.44 km/s towards an escape velocity of 11.2 km/s. Even an early Earth, with a 12-hour day, would only yield a gain of 0.88 km/s. In the following baseline assessments this component is therefore ignored.

Here it is suggested that a figure of merit of orbital/maneuver energy costs for IHZ evaluations could be constructed by aggregating the following measures:

i) The body-to-space Delta-v budget.

ii) Orbit-to-orbit Hohmann transfer Delta-v distributions.

iii) Mean Delta-vs for lower energy transfer via specified gravity assists or to reach specified Lagrange points to enable ultra-low energy weak stability transfers.

Given the challenges involved in systematically assessing iii), the measures of i) and ii) are of more practical use to build a first-order figure of merit for the IHZ. However, while the body-to-space Delta-v (by default the escape velocity) is readily computed for bodies across a planetary system and could be summed for a global figure of merit for the IHZ, the orbit-to-orbit Delta-v measures are a little more complicated. Hohmann transfer Delta-vs can be calculated for the idealized case of transfer between two coplanar circular orbits, with 2 applied Delta-vs (initiating transfer and then matching target orbit). The required Delta-vs are computed in the standard way using the vis viva equation: $v^2 = GM_* \left( \frac{2}{r} - \frac{1}{a} \right)$. But there are a variety of ways to aggregate these transfer costs to guide an assessment of the IHZ.

For example, optimal planet-to-planet transfer Delta-v's could be summed for all body location pairs in a system, but an interplanetary species might utilize a wide range of orbits, for habitats, staging, or other needs. Furthermore, not all body location pairs are necessarily important for the IHZ, especially if some planets do not provide easy resource access (e.g. moonless giants). Alternatively, the range in orbital radii that is accessible with a fixed Delta-v budget could be computed from any radius or from each body's orbital radius, under the assumption that those locations will play a major role in transfer operations.

In Section 3 below a simplifying approach is used to develop an IHZ figure of merit where Delta-v's are only computed between a point of origin (e.g., Earth) and specific locations (e.g. other planetary bodies) for all interplanetary transfers.

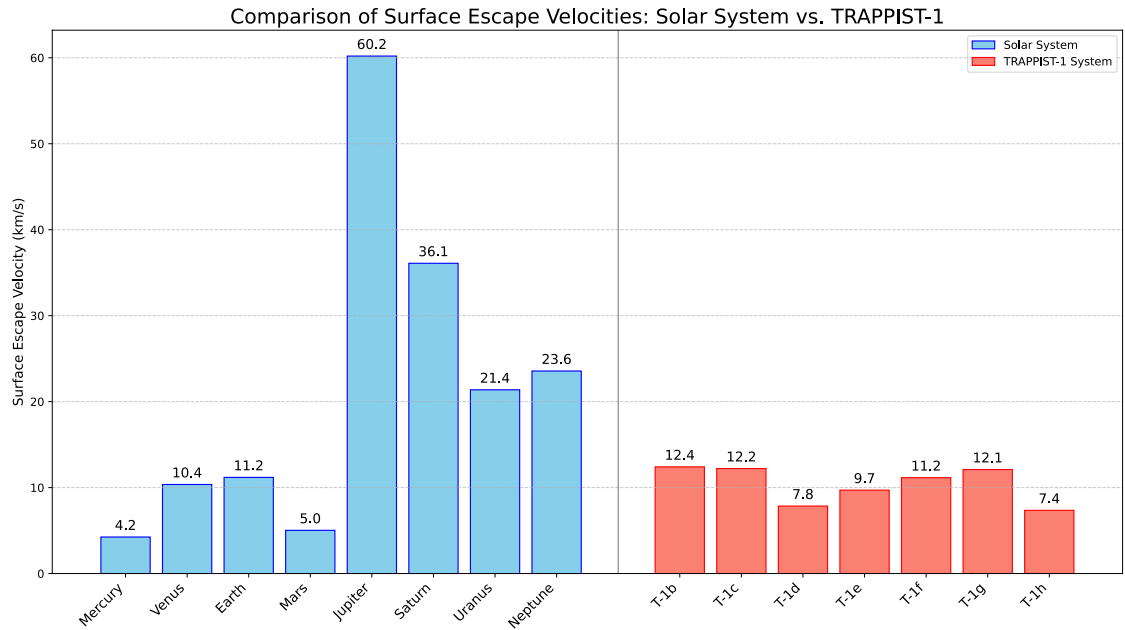

*Figure 3: Surface escape velocities of planetary bodies in the solar system (left panel) and estimates for the Trappist-1 exoplanetary system (right panel).*

## 2.4 Material resources

The interplanetary habitability of a system must be highly dependent on the availability of material resources, just as habitability on an originating planetary surface is dependent on local material resources, such as compounds supporting metabolism and growth. The IHZ will have a dependency on natural resources on planetary bodies and asteroidal bodies that can be utilized both by the mechanisms of interplanetary habitation (e.g. technology for transport or settlement) and biological life.

The simplest starting point for quantifying the resources of a given body is to assign a resource weighting $\alpha_{resource}$ that attempts to capture the overall availability of materials, including their richness (i.e., elemental variety, compound diversity, abundances), and accessibility (i.e., surface or near surface, concentrations, and physical or chemical extractability). For example, in the solar system $\alpha_{resource}$ might be very roughly estimated for major locations as (also see Section 4 below): Mercury: 0.4, Venus: 0.3, Earth: 0.5, Moon: 0.6, Mars: 0.7, Asteroids: 1.0, Jovian system (i.e. moons): 0.8, Saturnian system (i.e. moons): 0.6, Uranian system (i.e. moons): 0.3, Neptunian system (i.e. moons): 0.3. These relative weightings are not rigorous but reflect broad estimates of composition (including key metals and water), and accessibility (i.e. placement and concentration of resources, where for example Earth's oceans reduce some accessibility compared to the Moon or Mars).

It is widely acknowledged that in our solar system asteroids represent one of the most accessible and potentially useful resources for an interplanetary economy and its infrastructure (e.g., Trigo-Rodriguez et al. 2025 and references therein). For the IHZ of a system the following principal characteristics of asteroids are likely to be important:

i) Total mass and mass function of asteroid bodies.
ii) Composition function of asteroid bodies (elemental abundances and chemical abundances).
iii) Accessibility determined by orbital properties and material state and depth of resources in asteroid bodies.

Asteroid bodies may occupy a wide range of orbital configurations that can increase their accessibility but also complicate the calculation of orbital transfer costs across a system. For example, in our solar system, Sloan Digital Sky Survey (SDSS) data has provided detailed information on the distributions of asteroids (e.g. main belt objects, DeMeo and Carry 2013, 2014) and ongoing surveys are adding increasing amounts of data (e.g., initial results from the Rubin Observatory, Greenstreet et al. 2026).

Figures 4 presents the solar system's semi-major axis distribution of asteroids captured in a dataset derived from the JPL Small-Body Database which contains all known asteroids and comets (downloadable from Kaggle as collated by M. S. Hossain: see Hossain and Zabed 2023). Figure 5 presents the computed perihelion and aphelion distances of the same population.

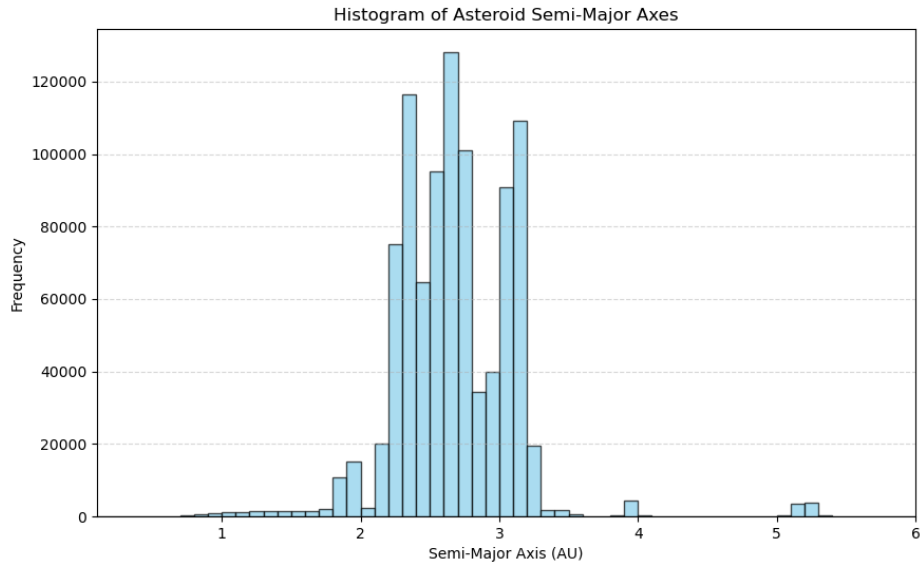

*Figure 4: The occurrence number of asteroids in 0.1 AU bins across the range of 0.1 to 6 AU in the solar system, from a JPL-SPD derived dataset of 900,000 identified objects.*

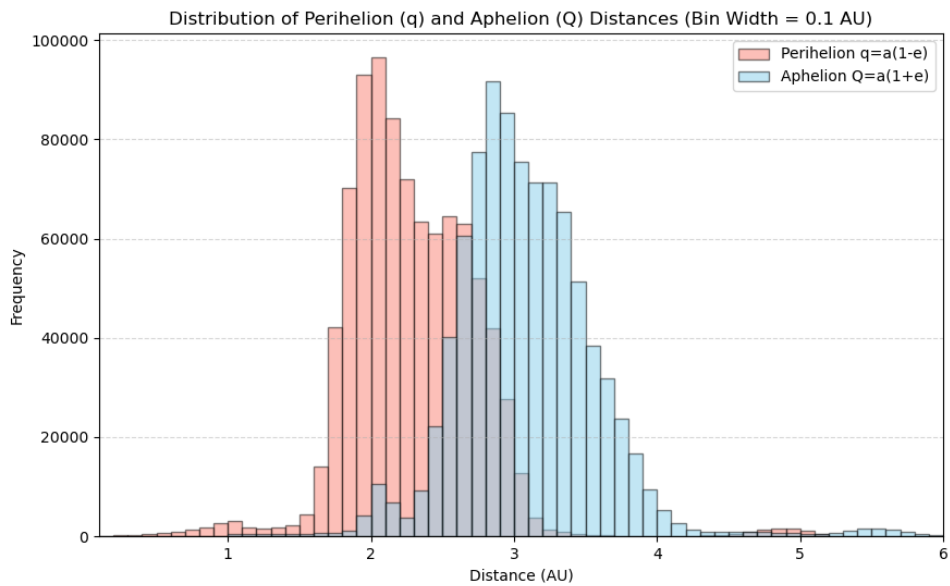

*Figure 5: the distribution of perihelion and aphelion distances for 900,000 asteroids, illustrating the range of orbital locations providing access to these resources (modulo orbital inclinations, which can be significant).*

As is shown in Figures 4 and 5, the orbital range for most known asteroid bodies in the solar system results in potential intercept radii of anywhere from ~1.5 AU to ~4 AU, with additional populations extending into the inner and outer solar system. However, orbital inclinations of more than 10⁴ asteroids exceed 10° and since transfers to such objects can incur a Delta-v cost of $\Delta v = 2v \sin\left(\frac{\Delta i}{2}\right)$, where $\Delta i$ is the required inclination change, such objects may be less accessible even if their orbits bring them closer to any transfer point of origin.

For planetary bodies in general there are of course other factors that can contribute to the effective resources of a location. For example, a measure of the classical habitability could be included. More habitable environments could presumably amplify the utility of material resources and be considered as a type of resource. In this initial work those additional factors are left for later study (see Discussion below).

## 3. IHZ metrics and functional forms

The classical CHZ has proven useful because its many contributing factors (including spin-orbit configurations, planet mass and composition, atmospheric structure and chemistry, and stellar properties leading to climate states) can be combined and captured in physical models to yield estimates of surface thermal conditions that can be treated as a relatively simple metric of habitability. By contrast, the IHZ as proposed here is a multi-modal evaluation of factors that may be entirely independent of each other (e.g., radiation risk in space may have no causal correlation with planetary resource richness).

One approach to constructing a simplified IHZ metric is to establish a normalized baseline function for each factor (each mode) and combine these into a single measure of relative interplanetary habitability. For example, planetary body resources can be characterized with a dimensionless relative weight $\alpha$ in some range $[0, x]$ normalized to the maximum value in a system. Steady-state (or time-averaged) radiation risk, $\lambda$, can be normalized by its maxima as a function of orbital radius within a range considered for the IHZ. Orbital transfer costs, to/from different orbital radii can be tied to a single location. For example, from/to the planetary body seen as most likely to originate life (i.e. that is within the CHZ) and normalized to the maximum transfer Delta-v (including surface-to-space costs) in the system.

A system-specific, radial relative habitability function $IHZ(r)$ incorporating these factors can then be written in the following general form:

$$IHZ(r) = \frac{P_*(r)}{P_{max}} + \frac{\alpha(r)}{\alpha_{max}} - \frac{\lambda_{radiation}(r)}{\lambda_{max}} - \frac{\Delta_{orbit}(r)}{\Delta_{max}}, \quad (1)$$

where $\Delta_{orbit}$ is a placeholder for the several potential options for calculating an energy/cost factor for orbital transfers and body-to-space Delta-v costs. Written this way the IHZ explicitly balances power and resources against risk and energy costs for interplanetary activity.

This might then be summed over a finite set of key locations, (e.g., planetary bodies) to arrive at a single figure of merit for a planetary system:

$$IHZ_{total} = \sum_i IHZ(r_i), \quad (2)$$

In Figure 6 an example of the application of these IHZ definitions is presented for the solar system. For full details of the parameterizations and values used see Sections 2.1-2.4 and 4. Solar system properties are assumed for power (and temperature dependency) and particle radiation risk, and resource weightings are chosen in a scheme broadly applicable to our current understanding of planetary, moon, and asteroid compositions (see above and section 2.4). Here the asteroid resources represent a maximum richness in the system of 1.0 (see also Section 4). Orbital and off/on transfer costs are determined only between the Earth and each other location and would therefore differ if alternate approaches were used (see above).

In this example the figure of merit for the entire system is $IHZ_{total} = -1.14$. This negative value appears to be determined largely by the risk and cost of the outer planets in the scenario set by the values of resource richness and the metric used for Delta-v costs.

For comparison, in Figure 7 an equivalent set of calculations are presented for the Trappist-1 exoplanetary system (see Section 4 for details), using Trappist-1e as the reference planet for all Delta-v transfers (i.e. all transfers computed between e and the other locations). It should be noted that an entirely hypothetical outer debris/asteroid disk is included in this model (see Section 4). Stellar particle radiation is also assumed to be significantly more severe than in our solar system (Garraffo et al. 2017). For Trappist-1 the figure of merit is $IHZ_{total} = -1.64$, or some 43% lower than for the solar system.

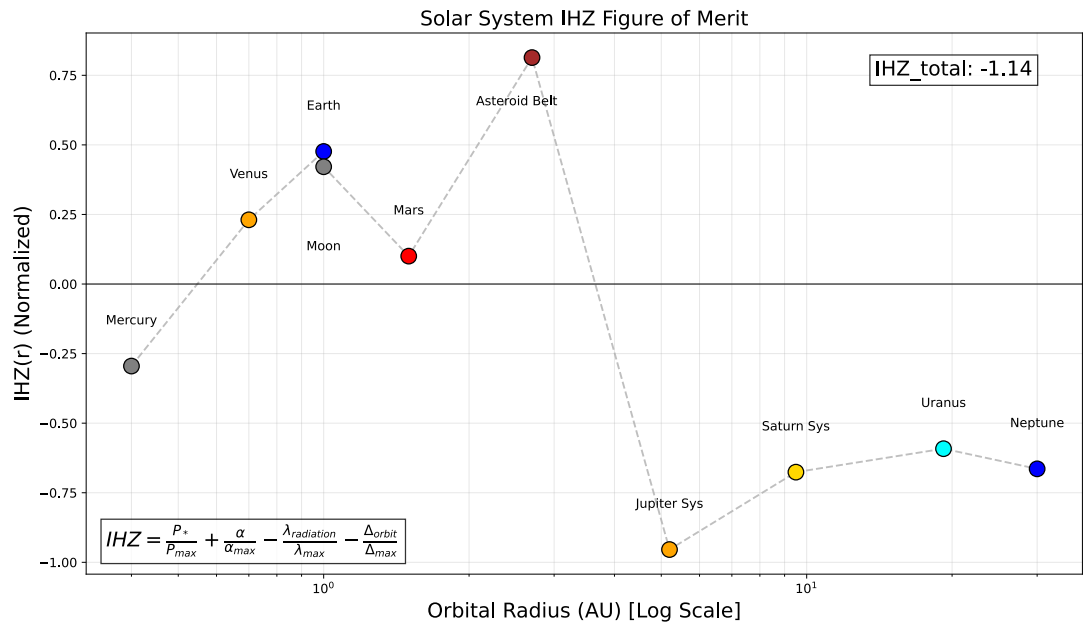

*Figure 6: An estimate of IHZ(r) is plotted for the solar system according to Equation 1 and the scheme described in the text.*

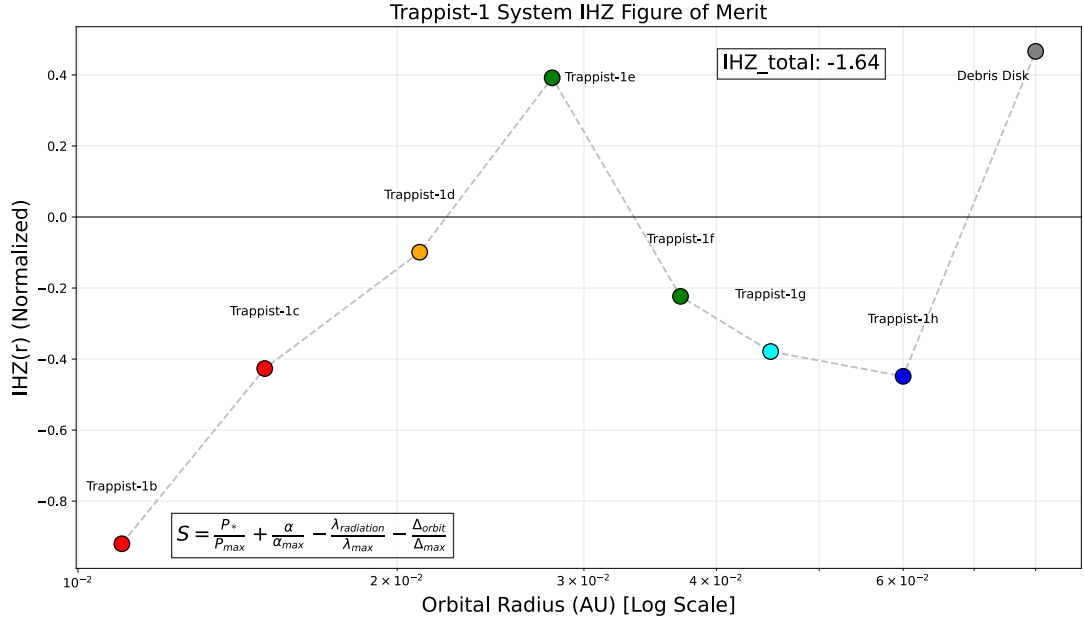

*Figure 7: An estimate of IHZ(r) is plotted for the Trappist-1 planetary system, with normalization to maximum values within this system.*

However, the normalization of the IHZ within a given system (Equation 1) means that the comparison of $IHZ(r)$ or $IHZ_{total}$ between different systems should be made carefully, since they are indirect metrics of whether one system is more "interplanetary habitable" than another. A more general option would be to normalize all factors relative to the solar system peak values, i.e.:

$$IHZ(r) = \frac{P_*(r)}{P_{max}^\odot} + \frac{\alpha(r)}{\alpha_{max}^\odot} - \frac{\lambda_{radiation}(r)}{\lambda_{max}^\odot} - \frac{\Delta_{orbit}(r)}{\Delta_{max}^\odot}, \qquad (3)$$

where ⊙ denotes values for the solar system. An example of this is shown in Figure 8 and can be compared directly to the results in Figure 6. In this case the summed $IHZ_{total}$ for Trappist-1 is approximately 10 × *lower* than for the solar system, a result that could be taken as implying that this exoplanetary system is very significantly less capable of harboring interplanetary life than the solar system. That figure of merit would be further reduced if the hypothetical outer debris disk were not included.

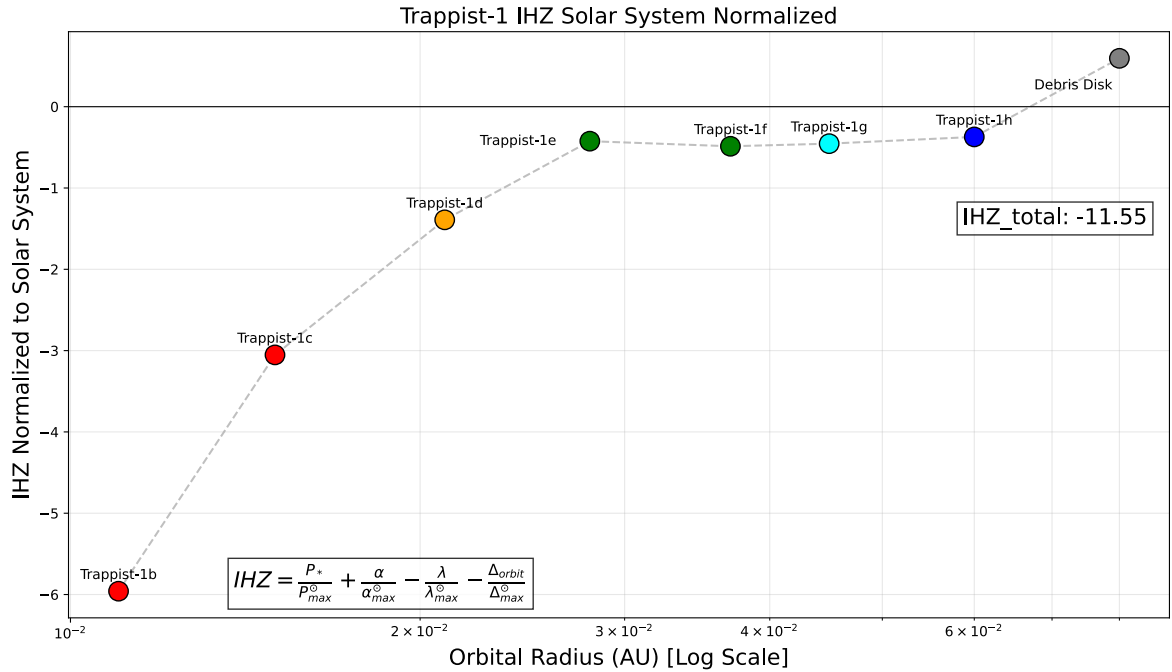

*Figure 8: The IHZ(r) function for the Trappist-1 planetary system with each factor normalized relative to the solar system peak values (Equation 3).*

A full interpretation of these results is beyond the scope of this present study and would require detailed investigation (and justification) of how the many parameters have been chosen (see also Section 4 below). Nonetheless, this appears to be a promising approach to quantifying the IHZ in a way that lends itself to comparison between different planetary systems.

## 4. An agent-based simulation approach to the IHZ:

A more comprehensive approach to evaluating the properties of an IHZ is to model the success of space-faring life in a planetary system as a dynamic process driven by the collective behavior of that life. Population dynamics can, in general, be described by sets of coupled equations. However, such systems of equations can be cumbersome and difficult to solve and can yield non-physical solutions. In the case of an IHZ, with multiple potential locations and options for movement and resource use, such equations are likely particularly complex. An alternative approach is via numerical, agent-based simulation, that seeks to model the actions of a population of individuals following an appropriate rule set within a modeled environment.

To test this approach, a highly simplified agent-based model was built for the solar system (initially, see Section 4.1 below for application to Trappist-1) to illustrate the model characteristics and to explore the interplay of the different IHZ factors. One of the major simplifications used was to limit available orbital movement to Hohmann transfers between discrete orbital locations/nodes defined by the mean orbital radii of planets, the Earth's Moon as accessed from/to the Earth, and a single, averaged orbital location for the main belt asteroids. The Delta-v cost of planetary body surface operations is also included for all the major bodies, both on arrival (descent costs) and departure (escape velocity costs). For the asteroids this cost is set to a uniform minimum of 0.1 km/s (c.f. known values for Vesta, Ceres and smaller bodies). For the giant planets (e.g., Jupiter) a uniform cost equal to the surface (the top of visible atmosphere) escape velocity is very conservatively assumed to exist for operations at any inner moons.

The model starts with 1000 agents that perform actions during each 0.5-year timestep of the simulation. Those actions are: decision-making and potential movement, resource acquisition, replication, or death. The decision to move location is based on the calculation of a potential gain versus cost/risk score across all possible locations that are then ranked (see Appendix I for details), where gain or income is based on solar power and resource gathering, and cost/risk is based on Delta-v costs to new target locations, and radiation risks.

The initial conditions for a model run can be summarized as:

- **Population:** Starts with 1000 agents.

- **Originating location:** All agents originate on the surface of Earth (at 1.0 AU).

- **Resources:** Each individual agent starts with 80 resource units (a balance between over and under resourced, accounting for initial Earth escape Delta-v costs so that agents have options in the first timestep). These resources are conceived to be extractable/mineable material resources and power, i.e. elements and compounds for physical reproduction of agents and for fueling movement, plus solar power resource units. I.e., agents gain "income" through a combination of resource gathering and solar power input (see below).

- **Model map:** The solar system is represented by 10 specific locations ranging from the 8 major planets: Mercury (~0.4 AU) to Neptune (~30.0 AU), in addition to the Earth's Moon, and the asteroid belt, each with specific resource, solar power, and particle radiation rates. The asteroid belt is set at a single orbital radius location at 2.7 AU. And the Moon is treated as a distinct location coincident with the Earth's orbital semi-major axis but incurring landing/launch Delta-v costs.

Agent decisions and mechanics (e.g. resource gain/loss) are driven by the following factors (see Appendix I for full details):

- **Movement:** Travel costs in Delta-v for orbital changes are calculated using the usual vis-viva equation for Hohmann transfers, and resource (energy) cost is proportional to the Delta-v, with 3 resource units required for a Delta-v of 1 km/s. Arrival and departure Delta-v costs from planetary body surfaces are computed using solar system values, although asteroidal values are set to a uniform 0.1 km/s. However, the model does *not* account for synodic periods between objects, and transfers are allowed during any timestep. The model also does not account for time taken for any transfers, which are assumed to fit within the 0.5-year timestep to first order.

- **A resource economy:** Agents gain resources (e.g., mineral, material resources) and require solar power, and solar power also dictates the rate of resource acquisition (Appendix I):

i. **Mining/material resources and solar power:** Material resources are based on a location's planetary richness and accessibility, defined at the start of the simulation using arbitrary units, with baseline values of: Mercury – 0.4, Venus – 0.3, Earth – 0.5, Moon – 0.6, Mars – 0.7, Asteroids – 1.0, Jovian system – 0.8, Saturnian system – 0.6, Uranian system – 0.3, Neptunian system – 0.3 (see Section 3 above). In all cases, the inventory of resources is assumed to be steady state, i.e., there is no net depletion of resources across the solar system as time passes. The baseline total rate of material resource gathering for an agent is 8 units per timestep multiplied by the location specific richness factor. Baseline parameters are chosen to ensure a balance between the number of timesteps between movement decisions and the potential income of agents and their risk (including death, see below). It should be noted (see Discussion) that additional factors might contribute to what properties would count as resources, such as the traditional habitability of a location, or other motivations for utilizing a body (e.g., settlement as a hedge against planetary risk).

ii. **Solar power:** Modeled physically using the inverse square law for stellar flux and a temperature-dependent efficiency curve ($\eta$) to capture the fact that photovoltaic systems lose efficiency at higher temperature (see Section 2.1 above). The operational maximum is assumed to occur at a temperature of 298K ($25°$ Celsius). The available solar power contributes to the net resource income of agents per timestep (see above), and therefore also to the ability to reproduce based on the total resource units of an agent (see below).

- **A risk model and attrition:** The probability of an agent's death in a timestep is determined by the presence or absence of stored resources (with zero resources an agent dies automatically) and by comparing a random variable to a normalized total radiation exposure, which is the sum of:

- **Solar particle radiation:** Assumed to follow a $1/r^2$ drop-off with orbital distance (see above).
- **Cosmic particle radiation:** Increases linearly with orbital distance (as $0.015r$, see above and Appendix I).
- **Local radiation features:** Additional fixed penalties for local radiation environments. At Mercury (a 0.25 penalty), Jupiter (a 0.4 penalty factor), and Saturn (a 0.1 penalty factor) in this model.

The agents act as boundedly rational utility maximizers (i.e., making "good enough" choices). At every timestep they evaluate potential destinations by weighing projected reward or income (solar power plus mining resources) against travel costs (depletion of agent stored resources) and radiation risks, plus a random "roll of the dice" factor that can shift a decision back or forth across the threshold of risk versus reward (Appendix I).

The population growth dynamics is governed by successful agents that replicate when resource thresholds are met (>150 units in this case, compared to the initial conditions of 80 units per agent, to ensure sufficient ability to leave the Earth during early timesteps). Replication costs 100 units. Each timestep also incurs a base survival resource cost of 5 units per agent. As described above, attrition in each timestep is decided by an agent's inventory of resource units and a randomly drawn radiation death based on the risk at that location.

In the test models the total solar system agent population is capped if it exceeds 100,000 agents. In this case, during a time step, the total population is culled randomly and reduced back to 100,000. This is done to avoid computational overhead and cut-off exponential growth that would otherwise overwhelm the model and require the modification of the availability of resources (see above), which would require more elaborate parameterization of the coupled systems.

Several informative scenarios are seen in differently configured model runs. First, if the model is run using the baseline resource values (see above) it appears that the agent population remains located on the Earth for at least 100 years. In this case no agents make the decision to relocate elsewhere.

If the Moon's resource weighting is raised to ~0.75 (i.e. ~50% higher than Earth's default) a majority of model runs result in agents beginning to sustain themselves at the Moon within 100 years. In our own hypothetical future, this enhanced value attached to lunar resources could arise from the needs of a larger space economy or availability of a unique lunar resource such as $^3$He.

If, in addition, Mars's resource weighting is further set to ~1.5 (300% higher than Earth's) a majority of model runs result in agents building sustained populations at both the Moon and Mars after about 50 years. If the asteroid belt resource weighting is then set to ~1.7 (340% higher than Earth's) an agent population will also establish itself at that location within 100 years. In this scenario the Earth, Moon, Mars, and asteroid belt all come to sustain an agent population, with the precise numbers varying run-to-run.

In Figure 9 the results are presented of a single model run with these modified resource weightings. In this instance the agent population remains at the Earth and grows over time until around 50 years when the first successful relocations begin, first to Mars and then to the asteroid belt. Later, at 70-80 years there is a small population beginning to grow at the Moon. This scenario also demonstrates that the Earth-located population of space-faring agents must first grow sufficiently to ensure that the decision threshold to relocate is met by at least some of the agents.

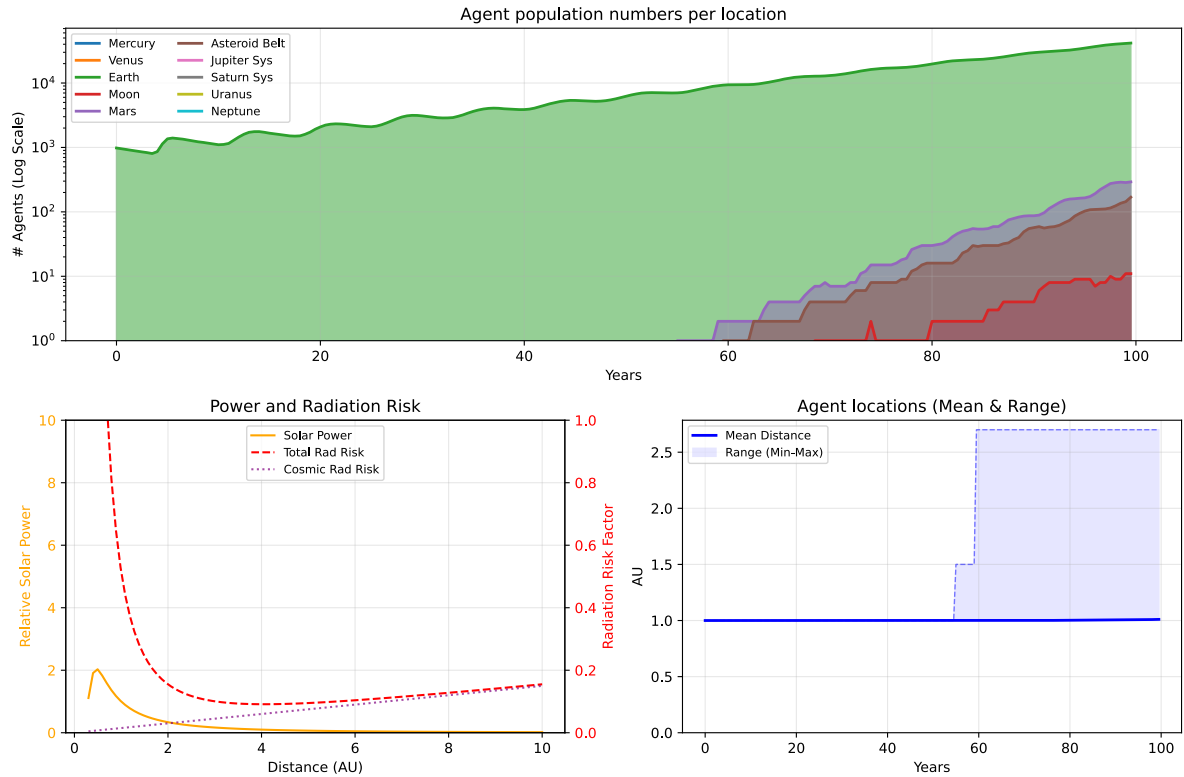

*Figure 9: The results of an agent simulation using resource weightings for the Earth, Moon, Mars, and asteroid belt selected to ensure outward growth of the population over time. Upper panel: the number of agents is plotted versus time for all locations with non-zero populations, areas under curves are shaded for visual clarity. Periodic variations in these curves reflects growth and attrition rates as resources are acquired and used over time, with growth occurring when a threshold is reached. Lower left panel: left y-axis – the relative solar power ($\eta \cdot f_*$) is plotted (solid curve) as a function of orbital radius. Right y-axis – the cosmic radiation risk is plotted as a function of orbital radius (dotted curve) and the total radiation risk (solar + cosmic) is plotted (dashed curve). Lower right panel: the mean orbital radius (solid curve) of all agents is plotted versus time, along with the orbital range spanned by all agents as a function of time (filled area).*

To further illustrate this dynamic, in Figure 10 the same model is run but with a modified/boosted lunar resource weighting of 0.9. This has a quite dramatic effect on the model history. The increased opportunity of lunar resources ensures an earlier migration from the Earth to the Moon around at around 20 years, followed by Mars at around 30 years, and the asteroid belt at closer to 70 years. I.e., the increased resource weighting of the Moon encourages migration to Mars ahead of the asteroid belt.

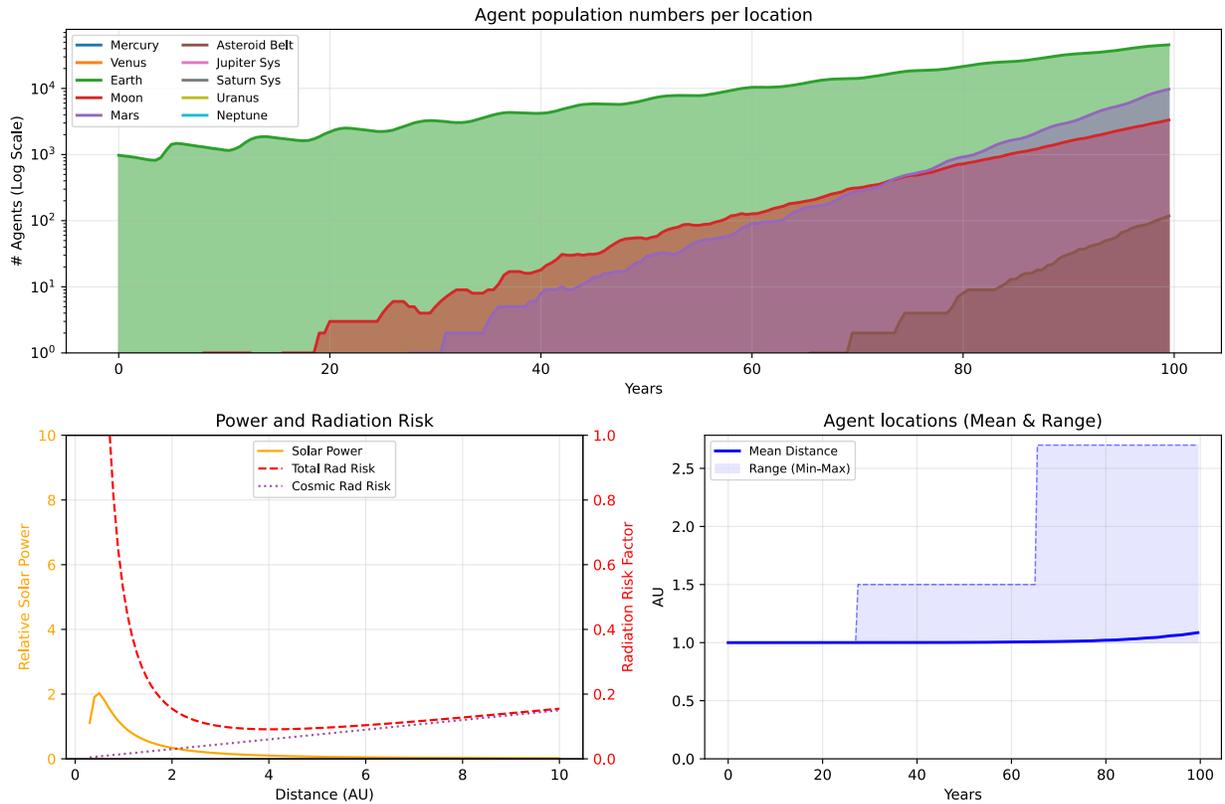

*Figure 10: Panel contents follow Figure 9. Here the lunar resource weighting has been increased to 0.9 to illustrate the impact on agent behavior at the Moon and on transfers to Mars that occur earlier in this model.*

In Figure 11 this same model is used but with an increased cosmic radiation risk function, boosting the slope to 0.05 compared to 0.015, thereby significantly raising the risk of attrition at larger heliocentric distances. This scenario might arise during a period of low solar activity and reduction of the heliospheric shield across the solar system. In this case the transfer of agents to Mars and the asteroid belt is significantly delayed by several decades and is attributable to the increased risk due to cosmic radiation both affecting the agents' risk/benefit decision making (since they have information about the risk parameters) and the attrition rates at those locations.

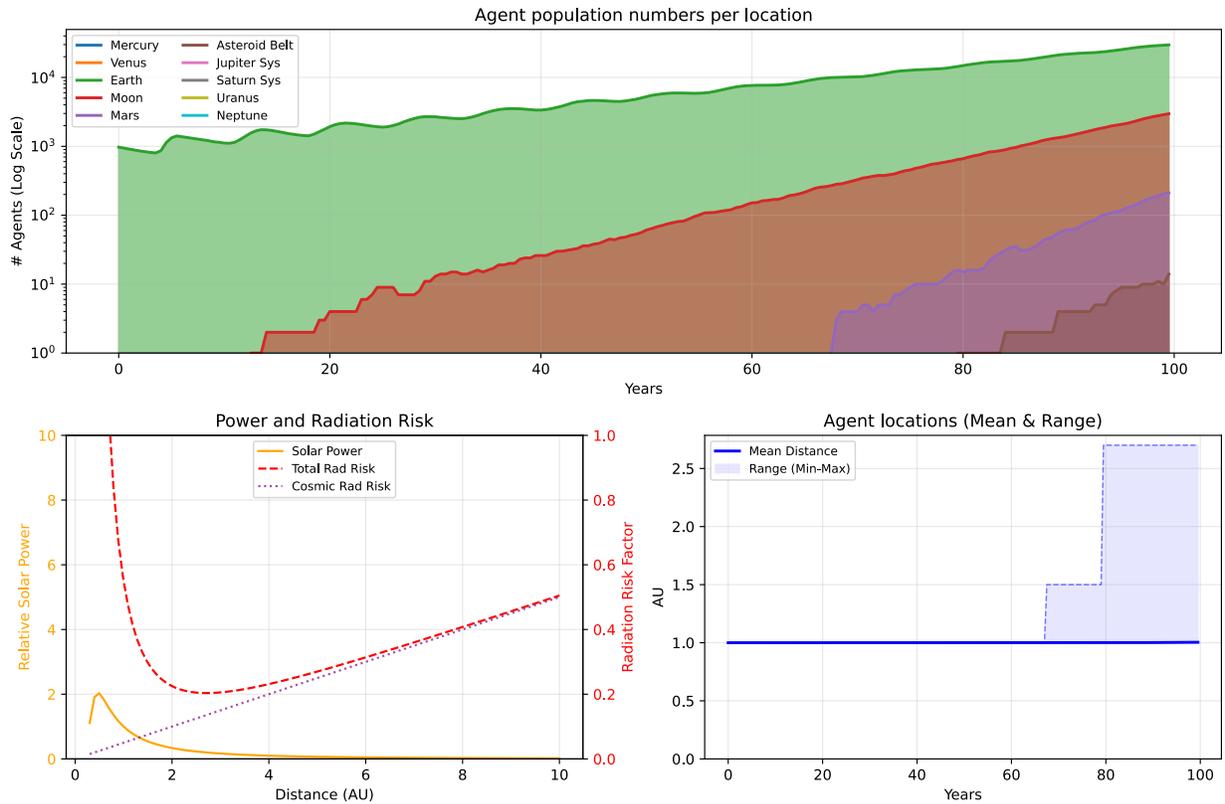

*Figure 11: In this model the cosmic radiation (GEP) risk has a higher slope (0.05 Versus 0.015 in Figures 9,10) that might be commensurate with a prolonged period of decreased solar activity leading to reduced heliospheric protection across the solar system (all other parameters are the same as the model shown in Figure 10). The IHZ dynamics are strongly affected, with a significant delay in agents reaching Mars and the asteroid belt*

To further explore the impact of resource weightings on population dynamics, a parameter sweep was performed on the resource weighting of each location in the model to establish what resource richness would cause any given location to become the eventual host to all agents. For each location the resource richness is varied from 0.0 to a maximum value of 30.0 (using an efficient iterative median search) and a full run of the agent model across the system is performed until the population locations have stabilized (typically within 100-200 years). The weightings of other locations are held at the default or baseline values (see above). The resource richness where >99% of the population stably occupies that location is recorded and shown in Figure 12. All other model parameters are also held fixed at the baseline values described previously. In this analysis a lower resource richness for occupation can be thought of as a higher overall score. Consequently, the ranking of locations in this model is: Earth, Moon, Mars, Asteroid belt, followed by Venus, and Mercury, while Jupiter – with high Delta-v costs and additional radiation risk – is least favored, and the outer planet systems are weighted similarly to Mercury.

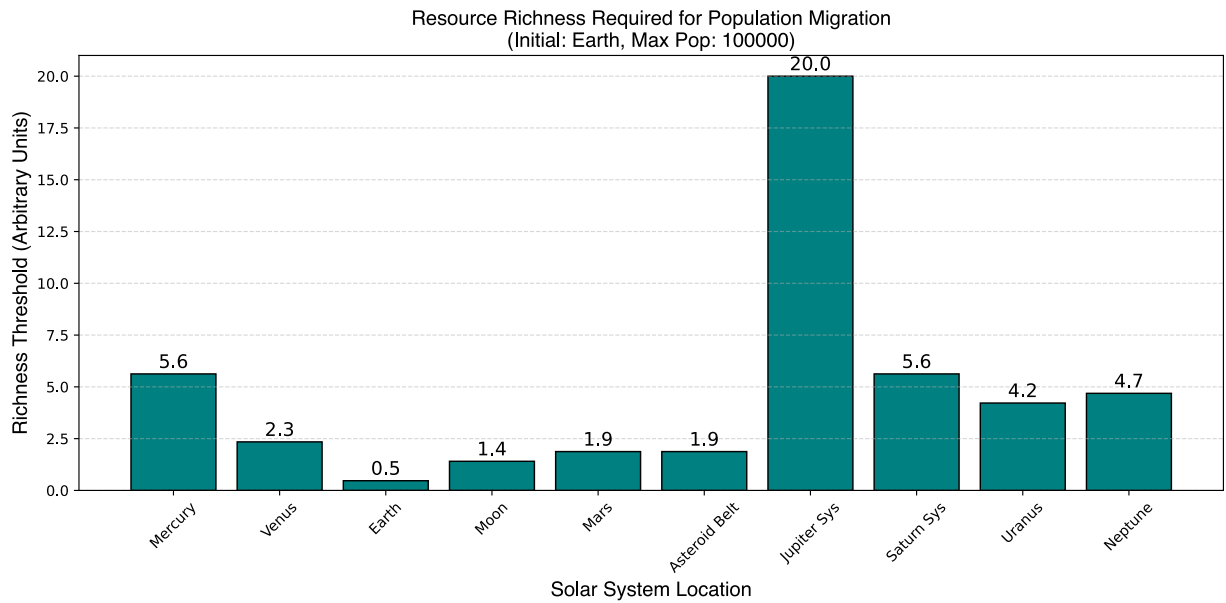

*Figure 12*: *The results are shown of a parameter sweep to determine the resource richness of each location in the model necessary to ensure that >99% of agents eventually migrate to that location, if all other resource weightings and parameters are held at their default values.*

The implications of these results are discussed further below (see Discussion). It is noted here that the resource weighting for Mars or the asteroid belt would need to be a factor ~4 × larger than Earth's for these locations to become the primary hub for agents.

### 4.1 Comparison of the Solar System and Trappist-1

The agent-based model is modified here to replace the solar system parameters with those of the compact Trappist-1 exoplanetary system, using published planetary orbital parameters, estimated masses and radii (used to estimate surface escape velocities), and stellar properties (e.g., Agol et al. 2021). In addition, an outer debris disk or putative asteroid belt is added as a location centered on an orbital radius of 0.08 AU. While there is currently no observational evidence for such a disk (Marino et al. 2020), models of planet formation do allow for the possibility. Here it is included to examine the consequences for compact planetary systems in general, using Trappist-1 as an archetype.

In this small system the effect of transfer orbit Delta-vs on agent decisions is greatly reduced compared to the solar system, while planetary escape velocities dominate the movement dynamics (these are estimated for the planets as: b: 12.2 km/s, c: 12.0 km/s, d: 7.8 km/s, e: 9.6 km/s, f: 11.0 km/s, g: 11.9 km/s, h: 7.3 km/s based on published masses and radii, Agol et al. 2021, and a notional 0.1 km/s is assumed for resources in the hypothetical debris disk). The initial agent population is placed at Trappist-1e as the planet generally thought to lie inside of most estimates of the CHZ orbital range (Lincowski et al. 2018). The stellar particle radiation is assumed to be high in the model. Studies of the potential Trappist-1 stellar wind magnetic environment and SCP events indicate that all planets may experience stellar wind pressures $10^3$-$10^5$ times higher than in our solar system and frequent flare events (Garraffo et al. 2017; Roettenbacher and Kane 2017). The baseline normalization of the inverse-square parameterized stellar particle flux is therefore weighted accordingly. In addition, the inner worlds b, c, d, e, g are assigned an inherent stellar radiation risk (c.f. Jupiter in solar system model) to further account for stellar flare activity. Resource weightings are obviously guesswork at this time, but for the sake of establishing baseline agent behavior are set to values commensurate with (or greater than) those of the solar system and in line with current estimates of the properties of these planets (i.e., rocky, icy, gas rich, etc.): b: 0.4, c: 0.4, d: 0.6, e: 0.9, f: 0.8, g: 0.7, h: 0.6, with the asteroid/debris disk set at a baseline minimum of 1.0.

Under these conditions the agent dynamics in the Trappist-1 model runs are notably different than for the solar system. In fact, the interplanetary agent population is vulnerable to extinction, as seen in Figure 13. By ~45 years all agents are gone, with the longest-lived population being those in the hypothetical debris/asteroid belt and only a modest amount of migration within the first 10 years to some of the other planets.

If stellar radiation risks are reduced by 50%, the situation is significantly altered (Figure 14). A larger population of agents accumulates across the outer planets for ~15 years before becoming extinct. However, an exponentially growing population is established at the debris/asteroid belt that persists to at least 100 years. In fact, the extinction of agents at the planetary locations appears to be mostly driven by their decisions to transfer to the debris/asteroid belt rather than because of radiation attrition. Without this resource location, for which there is currently no observational evidence, the Trappist-1 system would appear to be a poor environment for an interplanetary species.

It is noted that in this model an interplanetary capable population that remains at its point of origin (planet e) is not given any advantage (e.g. no assumed prior infrastructure or protection against particle radiation). There are of course many parameters that can be adjusted to explore the behavior of the system and a full exploration of these is beyond the scope of the present study. The implications of these results are explored in more detail in the Discussion below.

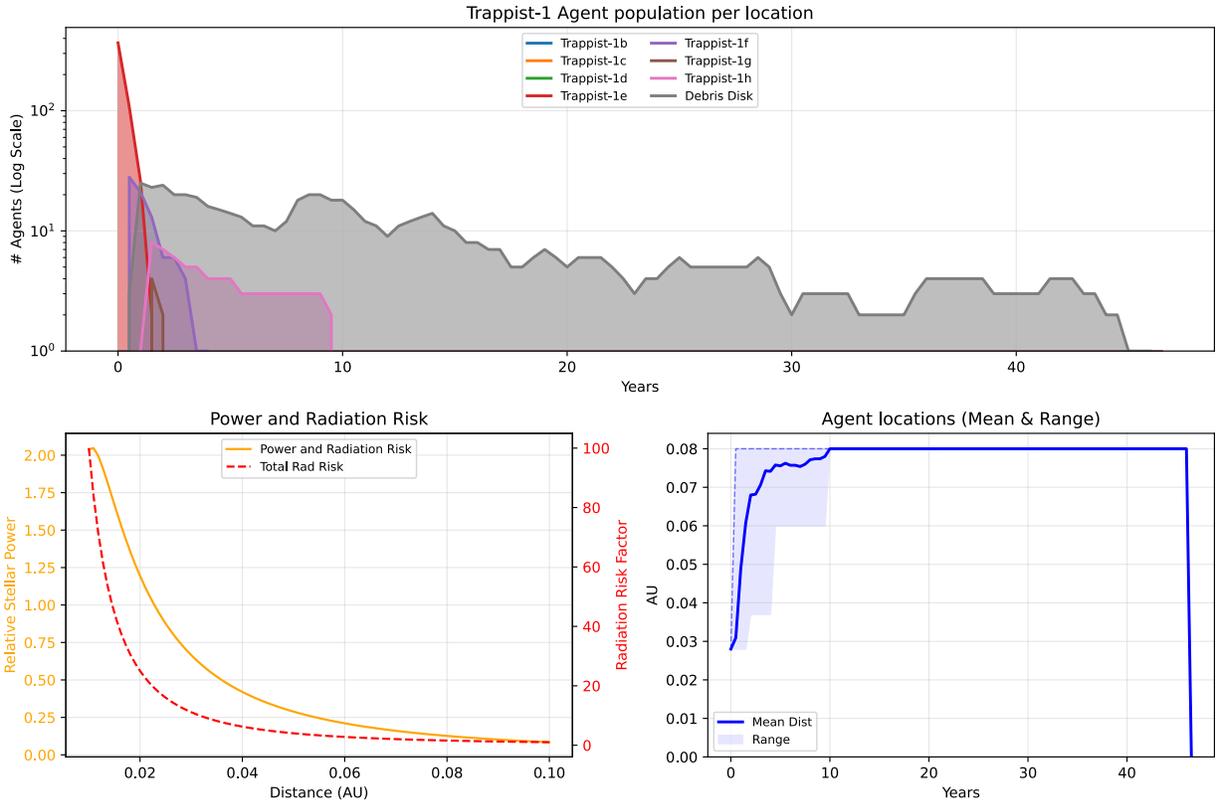

Figure 13: The results of an agent-based simulation of IHZ behavior for the Trappist-1 exoplanetary system. Default parameters are described in the text. Lower-left panel: only the total particle radiation risk is plotted, dominated by the stellar component and renormalized according to Equation 1. Despite early migration the agent population struggles to avoid extinction.

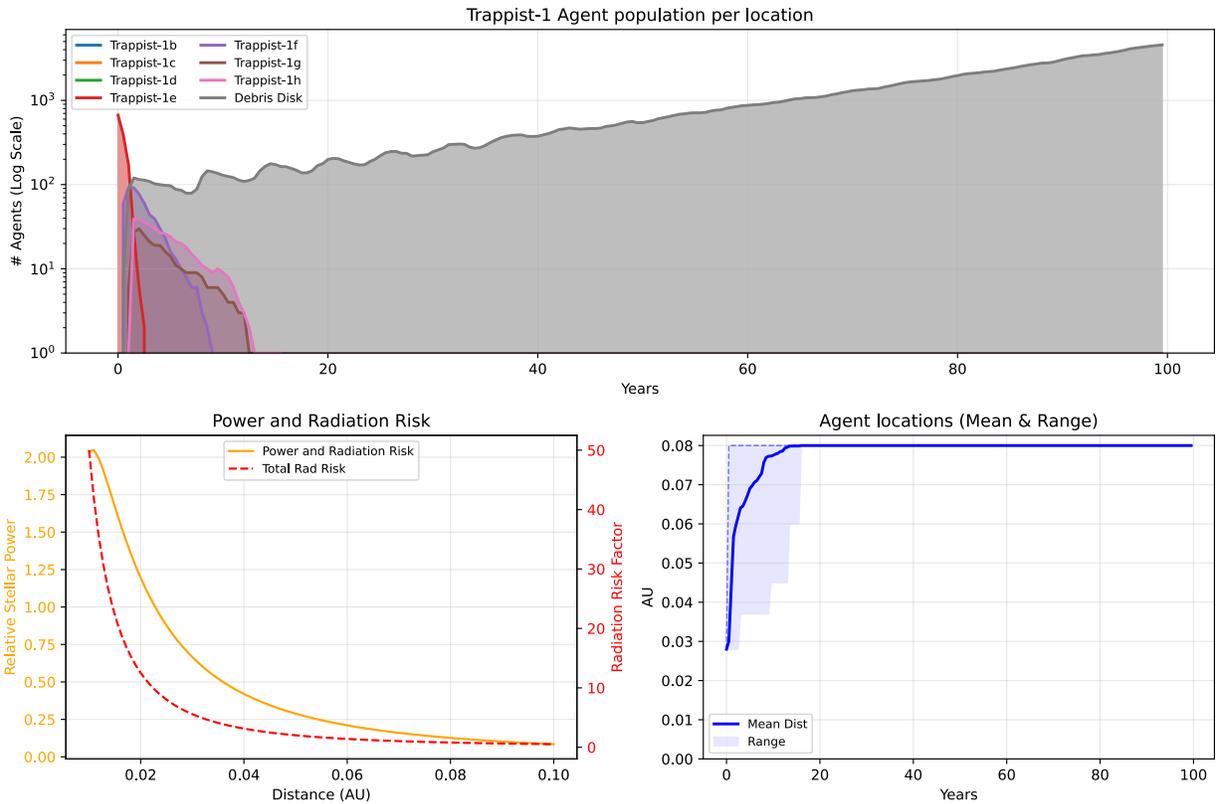

*Figure 14: The Trappist-1 model (Figure 13) with stellar particle radiation risk levels reduced by 50% from their baseline (note that the risk curve – lower-left panel – is renormalized following Equation 1). In this scenario a robust population is more likely to be established in the hypothetical debris/asteroid disk.*

## 5. Discussion

The concept of an interplanetary habitable zone (IHZ) is a very natural addition to traditional concepts of habitable zones and habitability in general. If life breaks free from its planetary origin points and establishes itself across other parts of a planetary system that represents a fundamental change in the equilibrium conditions for such life. This could include the establishment of new biospheres (and technospheres) on other bodies in the system that would yield potential biosignatures for detection. An interplanetary existence could also produce technosignatures that differ from those that might occur from species confined to a planetary point of origin. These techosignatures might include interplanetary communication systems, unexpected modification of other bodies (e.g., atmospheric compositions), and the formation of new technological structures with unique, observable, characteristics (e.g. in reflected or absorbed starlight or radiation emission).

The IHZ is also important to understand here in our own solar system. We are arguably in the process of becoming an interplanetary species, starting some 70 years ago and with accelerating growth today. But understanding the dynamics of our exploration and potential exploitation of the resources of the solar system requires a rigorous framework. The IHZ, and the type of agent-based simulation presented here could help provide such a framework.

Agent-based simulations of the solar system indicate the potential of the Moon, Mars, and asteroid belt as locations for resource-seeking, survival-maximizing entities. But also suggest that the value of these other locations (measured via the broad definition of resources used here) must be significant for an Earth-originating species to make the decision to migrate. Specifically, with the resource weightings used here, the Moon needs to represent at least a 50% more attractive proposition than the Earth for space-faring agents, Mars at least a 300% more attractive proposition, and the asteroid belt at least a 340% more valuable location. This is particularly acute for Mars or the asteroid belt if cosmic particle radiation becomes a larger risk due to solar activity minima. Furthermore, for the Moon to be the first site of migration it needs a resource weighting approximately 80% higher than the Earth for space-faring agents. This would also result in earlier agent migration to Mars.

Evaluating the resource weightings ($\alpha$) for our solar system or any other system is key to modeling the IHZ. In the present work this is done in a largely ad hoc fashion. Future work should consider further decomposing $\alpha$ into sub-factors that include measures of mineralogical/element abundance, utilization costs (including accessibility), and other values for an interplanetary species (e.g., use for fabricating structures, industrial use, propulsion needs, support of biospheres/habitats etc.)

The comparison of the results of the agent-based model of the solar system and the Trappist-1 system is a clear example of how two different planetary system architectures and parent stellar types might impact the IHZ. Given the very large uncertainty (arguably unconstrained) on characteristics such as resource richness/availability, any interpretation of the model outcomes should be made cautiously. Nonetheless, it is interesting to note that if a system like Trappist-1 has significant stellar radiation risks and does not have a reasonably low Delta-v accessible resource such as an asteroid/debris disk (or perhaps a resource-rich moon around the point of origin) then the IHZ (as modeled here) could be very restrictive. This is somewhat counterintuitive given the small planet-to-planet Delta-v costs in this system.

The results presented here indicate that in a compact system like Trappist-1 a space-faring species might only persist if it, i) is capable of harvesting resources at a high rate, and ii) succeeds in migrating to a larger orbital radius, such as a resource rich, low on/off body Delta-v asteroid belt, or iii) significantly reduces vulnerability to stellar particle radiation and flares. It is however possible that massive satellites/moons do exist around the major planets (Dey and Raymond 2025), which could modify these conclusions. Otherwise, there may be little impetus for species to shift to an interplanetary existence.

It is important to note that the IHZ as presented here does not incorporate any explicit measures of the traditional habitability of any object, or other reasons why an object is favored. Specifically, the factors considered in Section 3 and the agent-based simulations used here (Section 4) do not include any information about suitability for settlement of a population or other motivating factors. Mars, for instance, does not gain weighting as a potential location to support a large biosphere or population of an otherwise space-faring species. That weighting could, for example, depend on a species' interests in hedging against risk by settling other planetary locations. However, Figure 12 indicates that Mars would need to be a factor of ~4 times more favored/weighted over its baseline resource value to become the primary location for the population of agents in this model of the IHZ. Presumably if a species faced a very high risk of existential threat on Earth that factor of 4 could be met.

The agents in the simulations presented here do not make strategic plans. For instance, there is no mechanism in the model for the agent population to assess the risk versus reward of first migrating to the Moon or Mars before reaching the asteroid belt, or any other multi-step strategy. That capability could be incorporated in future models and might be particularly relevant for evaluating the potential for a bona-fide space economy in our own solar system. Furthermore, the model agents have a single, shared set of goals that pertain to replication (driven by resources and solar power) and risk reduction (reducing attrition rates). Agents are not engaged in settlement building or spinning off other infrastructure that would increase their efficiency at a given location. They therefore represent the "purest" type of population that could sample an IHZ that is not altered by their presence (c.f. biospheres certainly modify the habitability of environments in the classical CHZ).

To reiterate, the agent-based simulation here assumes a monoculture of agents. They have a shared set of goals and are not in competition with each other, nor do their efficiencies evolve with time (e.g., as new capabilities in movement or resource utilization emerge). A future exploration of these models could include the introduction of competition and a capacity to evolve in the face of that competition.

**Acknowledgements**: The author acknowledges the support of the NASA Planetary Science Division Astrobiology Program and the NASA Ames Research Center. The Python software used to generate the plots shown here, and to develop the agent-based simulations, was initiated utilizing Gemini Pro 2.5 to draft potential code that was then scrutinized, tested, and modified by the author.

## Appendix I: Agent decision mechanics

The agent decision-making process involves calculating a score (S) for each potential location $d$ (including the agent's present location $l$) and the agent moving (or remaining stationary) to the location with the highest S (if sufficient resources are available for the required Delta-v).

$$S_d = \left(5 \cdot I_{total}(d)\right) - C_{move}(l,d) - \left(50 \cdot R_{risk}(d)\right) + \epsilon$$

Where $I_{total}(d)$ is the total projected income at the potential destination, calculated as a sum of solar power income and resource income:

$$I_{total}(d) = \left(\eta(d) \cdot f_*(d) \cdot I_{base}\right) + \left(\alpha(d) \cdot \gamma_{gather}\right)$$

Where $I_{base}$ is a base solar power income constant, $\alpha(d)$ is the resource richness at location $d$, and $\gamma_{gather}$ is the base resource gathering rate (default of 8 units).

$\epsilon$ is a random noise term drawn from a normal distribution and added to ensure that agents decisions are not entirely uniform. This could be considered as a representation of agents having imperfect information or imperfect function at any time step, or a limited degree of self-determination.

The resource cost to move from location $l$ to $d$ is $C_{move}(l,d)$ and is calculated as:

$$C_{move}(l,d) = \Delta_v(l,d) \cdot C_{\Delta v}$$

Where $\Delta_v(l,d)$ is the total Delta-v required (escape, transit, landing) to move between locations, and $C_{\Delta v}$ is the resource unit cost per km/s of Delta-v (default of 3 units).

Finally, $R_{risk}(d)$ is the particle radiation risk at potential destination $d$:

$$R_{risk}(d) = \frac{0.5 \cdot \lambda_{star}}{r_d^2} + P_{rad}(d) + \left(\lambda_{cosmic} \cdot r_d\right)$$

Where $r_d$ is the distance of the potential destination from the host star, $\lambda_{star}$ is a factor setting the amplitude of stellar particle radiation, $P_{rad}$ is an additional location dependent radiation penalty for some locations only (e.g., Jovian system), and $\lambda_{cosmic}$ is the slope of the cosmic radiation function (assumed to be linear). All other numerical factors are chosen to ensure plausible behavior during each timestep.